\documentclass[reprint,superscriptaddress,showkeys]{revtex4-1}
\usepackage{verbatim}
\usepackage{color}
\usepackage{subfigure}
\usepackage[]{graphicx}
\usepackage{dcolumn}
\usepackage{bm}
\usepackage{blindtext}
\usepackage{floatrow}
\usepackage{makeidx}
\usepackage{epsfig}
\usepackage{hyperref}
\usepackage{array}
\usepackage{epstopdf}
\usepackage{braket}
\usepackage{multirow}
\usepackage{mathtools}
\usepackage{amsmath,amssymb,amsfonts}
\usepackage{fancyhdr}
\usepackage{multirow}
\newcommand{\RNum}[1]
{\uppercase\expandafter{\romannumeral #1\relax}}
\newcommand{\Rnum}[1]
{\lowercase\expandafter{\romannumeral #1\relax}}
\def\qe{Quantum ESPRESSO}

\makeindex
\begin{document}
\preprint{APS/123-QED}
\title{Interaction between U-shaped amyloid beta fibril \\ and semiconducting silicon nitride monolayer}
\author{Ashkan Shekaari}
\email{shekaari.theory@gmail.com}
\affiliation{Department of Physics, K. N. Toosi University of Technology, Tehran, Iran}
\author{Mahmoud Jafari}
\email{jafari@kntu.ac.ir}
\affiliation{Department of Physics, K. N. Toosi University of Technology, Tehran, Iran}
\date{\today}
\begin{abstract}
Motivated by some recent works showing the ability of semiconducting monolayers to disintegrate the structures of biological fibrils, we have applied molecular dynamics (MD) simulations in both classical and quantum regimes to investigate whether semiconducting Si$_3$N$_4$ monolayer has the same ability on interaction with U-shaped amyloid beta (A$\beta$) fibril. In agreement with the literature, we found that disintegration began from the last chain (E) due to the rather strong interaction between the monolayer and the fibril residues numbered from 17 to 28 on the very chain, also engaging the next chain (D) over time. As a result, the $\beta$-sheet-rich content of chain E considerably decreases on interaction with the monolayer, turning into other secondary-structure types including turn and coil, in accordance with experimental findings. Results endorse the view that semiconducting Si$_3$N$_4$ monolayer has the potential of destabilizing the structure and conformation of U-shaped amyloid beta fibrils.\\
\end{abstract}
\maketitle
\section{\label{sec:1}INTRODUCTION}
Misfolding and aggregation of proteins including Huntingtin, $\alpha$-synuclein, and amyloid beta have so far been revealed as the main causes of their associated neuro-degenerative diseases namely Huntington, Parkinson, and Alzheimer (AD), respectively~\cite{R1}. Such protein aggregations indeed interrupt the functioning of normal cells, then adversely affect memory and motor activities depending on the brain region in which they accumulate. AD is the most commonly occurring disease in the elderly leading to dementia, depression, and death~\cite{R2}. The consecutive key events at the molecular levels of AD are: (i) the folding of amyloid beta peptides with $\alpha$-helical structures into disordered conformations; (ii) the formation of oligomers of $\beta$-sheet structures; (iii) the aggregation and further growth of these oligomers by means of monomer addition; and (iv) the formation of mature fibrils. Both oligomers and mature fibrils contribute to AD neuro-toxicity~\cite{R3,R4}. The toxicity of the latter (fibrils) is due to dystrophic neurites being found around amyloid plaques; the fibrils also induce neuronal cell death and generate oxidative stress~\cite{R5}. Treating AD at molecular levels should then include removing amyloid plaques from the brain, destruction of oligomers and mature fibrils, and inhibition of amyloid peptide aggregation processes either through stabilizing the native conformation (helical structure) or via destabilizing the disordered structures. Indeed, destabilizing these fibrillar entities is necessary to make them soluble so that the brain organelles could accordingly remove them. 

So far, several molecular structures have been proposed as fibril inhibitors such as polyphenol molecules, retro-inverso peptides, D-peptides, molecular tweezers, N-methyl peptides, quinone derivatives, and polyphenols~\cite{R6,R7}. In this regard, nano-particles have also widely been proved to be capable of inhibiting amyloid fibrillation, being also applied as secondary-structure modulators of amyloid peptide aggregates. Nano-epigallocatechin-3-gallate has been shown to be both fibril inhibitor and buster~\cite{R8}; gold nano-particles, due to tunable surface chemistry, have been studied in this regard~\cite{R9,R10,R11,R12,R13,R14}; polyoxometalates inhibit A$\beta$-peptide self-aggregation as well as metal-ion-induced aggregation~\cite{R15}; carbon-based nano-materials, such as fullerenes, carbon nano-tubes, graphene, and graphene oxide, have also shown aggregate-inhibiting and fibril-busting features on interaction with amyloid peptides and mature fibrils~\cite{R16,R17,R18,R19}; graphene monolayers showed the ability of both penetrating into amyloid fibrils and dissolving them via $\pi-\pi$ stacking interaction between the aromatic surface of graphene and Phe (phenylalanine) residues; graphene oxide has also experimentally been shown to be effective in busting the fibrils and disintegrating them into single peptide fragments~\cite{R20}.

Motivated by the discovery of graphene, other two-dimensional (2D) nano-materials such as WS$_2$, MoS$_2$, silicon nitride, and phagraphene also attracted the ardor of scientists, showing many intriguing properties arising from their semi-conductivity and low-dimensionality~\cite{R22,R24}. It has been shown that WS$_2$ surfaces bind to amyloid structures, inhibit their aggregation, and destabilize the associated fibrils, being also useful in imaging them due to suitable absorption properties~\cite{R26}; MoS$_2$ surfaces and nano-tubes modulate amyloid-peptide aggregation, and induce conformational changes in both $\alpha$-helical and $\beta$-sheet structures of the peptides~\cite{R1}. 

Based on the two features of semi-conductivity and two-dimensionality, we here investigate whether silicon nitride (Si$_3$N$_4$) monolayer is capable of inducing similar fibril-busting impact on structure and conformation of U-shaped amyloid beta fibrils. Indeed, the bio-compatibility of the monolayer has recently been proved~\cite{R27}; therefore, studying its possible abilities to induce favorable conformational changes in amyloid beta fibrils for destabilizing and disintegrating them is the next natural step towards proposing it as an amyloid fibril buster. Our investigation is based on molecular dynamics (MD) simulations~\cite{h,man} in both classical and quantum regimes.
\begin{figure}[H]
	\centering
	\includegraphics[scale=0.8]{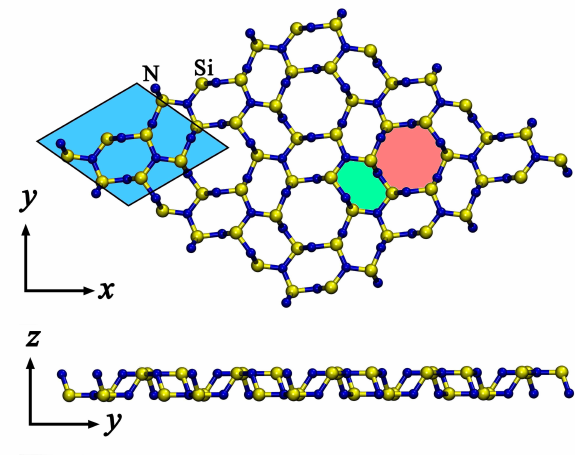}
	\caption{\label{fig:1}
		Atomic structure of silicon nitride (Si$_3$N$_4$) monolayer in $x-y$ and $y-z$ Cartesian planes. The blue parallelogram is the unit cell; the yellow and blue balls also denote Si and N atoms, respectively. As is seen, the atomic network is composed of 8- (green) and 12-atom (red) rings.}
\end{figure}
\section{\label{sec:2}Computational Details}
The following hardware/software setup has been adopted: initial atomic positions of U-shaped A$\beta$ fibril from RCSB Protein Data Bank (PDB) with entry code 2BEG~\cite{R28}; the NAMD Git-2021-07-13 computational code specified for Linux-x86\_64-multicore-CUDA~\cite{R29}; GeForce RTX 2080 SUPER graphics card; Debian-style Linux~\cite{R30} operating system; Open MPI v.3.1.6 for CPU parallelization; the 2019 update of CHARMM36~\cite{R31} force fields; TIP3P water model~\cite{R32}; the VMD (version 1.9.4a9)~\cite{R33} computational code for post-processing and visualization; two water boxes with dimensions $8.6\times 6.4\times 7.0$ and $18.1\times 16.5\times 6.6$ nm$^3$ (Fig.~\ref{fig:1}) respectively for simulating A$\beta$ fibril in water (abbreviated as A$\beta$-W) and A$\beta$ fibril plus the monolayer (ML) in water (A$\beta$-ML-W) under periodic boundary conditions with a unit-cell padding of $\sim 2$ nm to decouple periodic interactions; the system A$\beta$-W as the reference (control) trajectory to which the behavior of A$\beta$-ML-W is compared; electrostatic neutralization of the two systems via replacing five randomly-selected water molecules with the same number of Na$^+$ ions changing the total charge from -5e to $10^{-5}$e; the switching and cutoff distance values of 1.0 and 1.2 nm for truncating non-bonded van der Waals interactions, respectively; particle-mesh Ewald (PME)~\cite{R34} for long-range interactions; energy minimization simulation for each system by 100000 conjugate-gradient steps ($\simeq0.1$ ns); free-dynamics MD simulation for each system by 100 ns with the integration time-step of 1.0 fs, in NPT ensemble at 310 K and 1.01325 bar using Langevin forces with a damping constant of 2.5 ps$^{-1}$, along with the Nos{\'e}-Hoover Langevin piston pressure control; a dielectric constant value of 1.0 for A$\beta$-W (due to a very large amount of vacuum on nano-scale compared to A$\beta$-W atoms); and the hydrogen donor-acceptor distance and the angle cutoff values of about 3 \AA\ and 20$^\circ$ respectively, to identify hydrogen bonds.
\begin{figure}[H]
	\centering
	\includegraphics[scale=0.32]{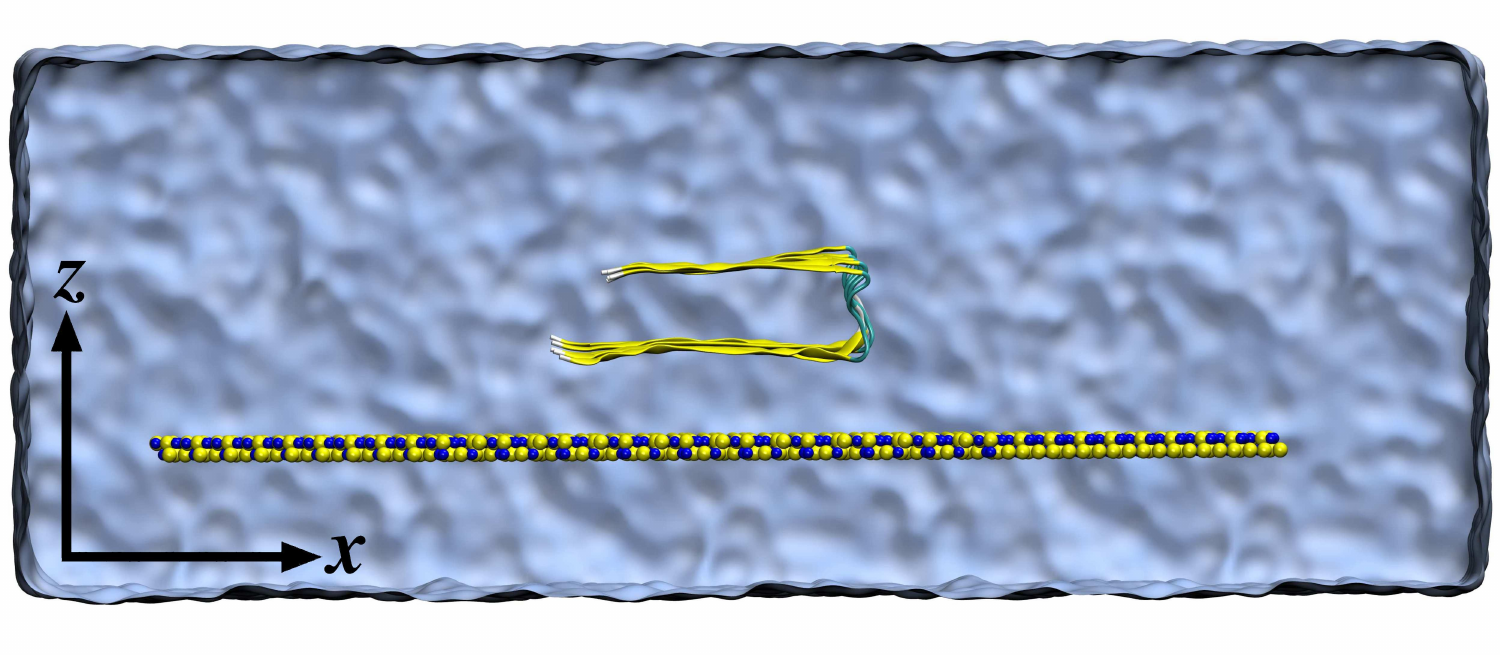}
	\caption{\label{fig:2}
		Side view of A$\beta$-ML-W. The fibril axis is along $y$ pointing outward.}
\end{figure}
Solvent-accessible surface area (SASA)~\cite{R35} was also calculated using the rolling-ball algorithm~\cite{R36} with a radius of 1.40 {\AA} for the probe sphere.

It should be noted that similar to every protein, the A$\beta$ fibril is composed of several domains; therefore, its initial orientation with respect to the monolayer could have a considerable impact on the results. According to Fig.~\ref{fig:1}, we have investigated a specific configuration in that the fibril axis is parallel to the surface of monolayer in order to come up with the largest contact area, having the maximum number of 15 amino acids on each of the five chains, between the fibril and monolayer. The overall distance between the fibril and the monolayer also increased on average from about 2.0 to 4.0 \AA\ during minimization. All the MD simulations were repeated two times with slightly different initial atomic configurations in order to check the reproducibility of the results.
\subsection{Force field of 2D-Si$_3$N$_4$}
To obtain the force field of 2D-Si$_3$N$_4$, we started with the values associated to the bulk state of $\beta$-Si$_3$N$_4$, reported by Wendel {\em{et al.}}~\cite{R37}, then calibrated them in a way that the dielectric constant of the monolayer calculated by NAMD (classical value) was fitted to its quantum mechanical analogue $\kappa\simeq2.370$.

To compute the dielectric constant quantum mechanically, we adopted the following parameters: a self-consistent, plane-wave, pseudo-potential approach at PBE-GGA~\cite{R39} level of density functional theory (DFT)~\cite{R40} as implemented in \qe~\cite{R38}; scalar-relativistic ultra-soft pseudo-potentials ($\mathtt{Si.pbe}$-$\mathtt{n}$-$\mathtt{rrkjus}$\_$\mathtt{psl.1.0.0.UPF}$ and $\mathtt{N.pbe}$-$\mathtt{n}$-$\mathtt{rrkjus}$\_$\mathtt{psl.1.0.0.UPF}$) generated by Rappe-Rabe-Kaxiras-Joannopoulos (RRKJ)~\cite{R41} pseudization method with nonlinear core correction~\cite{R42} to model the core electrons; the valence shells of N and Si described by $2s2p$ and $3s3p$ orbitals, respectively; optimized kinetic energy cutoff values of 90.0 and 360.0 Rydberg (Ry) respectively for the wavefunctions and the charge density; starting from the primitive cell of bulk $\beta$-Si$_3$N$_4$ containing 6 silicon and 8 nitrogen atoms on a hexagonal Bravais lattice under periodic boundary conditions with experimental lattice constant $a_0=7.820$ {\AA}; applying a vacuum space of $>15.0$ \AA\ along $z$ to produce the monolayer structure, as shown in Fig.~\ref{fig:2}; calculating the equilibrium (zero-pressure) lattice constant of the monolayer using Murnaghan's isothermal equation of state~\cite{R43}, leading to $a=8.2620$ {\AA}, being larger than $a_{0}$ by 5.65\%.

Car-Parrinello molecular dynamics (CPMD) simulations were carried out, including one minimization and one finite-temperature simulation at 310 K, both in absence ($E_1=0.0$) and presence ($E_2=16.0$ kcal.mol$^{-1}$.{\AA}$^{-1}$.e$^{-1}$) of the electric field applied along $+z$ to change the dipole moment $p$ of the monolayer. Each finite-temperature simulation has been preceded by an electronic minimization to bring the electronic wavefunctions on their ground states relative to the starting atomic configuration. The two MD setups associated to $E_1$ and $E_2$ were minimized for 100 damped-dynamics steps ($\simeq$ 0.012 ps) with the time-step $\mathrm{\Delta} t=0.120$ fs, and with electron damping value of 0.10 ($=$ damping frequency times $\mathrm{\Delta} t$). 

The following parameters were also adopted as well: the fictitious electron mass of about 1000 a.u. (1000 times the mass of electron) in the CP Lagrangian to guarantee the validity of the adiabatic approximation~\cite{R44}; a mass cutoff value of 2.50 Ry for Fourier acceleration effective mass to keep the quality of simulations from adverse effects and to minimize the electron drag effect; starting from the same minimized ionic degrees of freedom for all CPMD simulations for which the total force exerted on each atom decreased to $<10^{-4}$ eV.{\AA}$^{-1}$; a propagation time of about 0.360 ps ($\simeq$ 3000 Verlet steps); a preconditioning scheme~\cite{R45} to accelerate electronic equations of motion; ignoring at least the first 0.120 ps ($\simeq$ 1000 steps) of the simulations for thermalization and reliable statistical averaging; and taking into account both electronic and ionic contributions in estimating the average value of total electric dipole moment. 

The dielectric constant was calculated according to the relation
\begin{align}
	\label{eq:diel}
	\kappa=1+\frac{\mathrm{\Delta} p}{\epsilon_{0}E\mathrm{\mathrm{\Omega}}},
\end{align}
where $\mathrm{\Delta} p=|p(E_1)-p(E_2)|$, $\epsilon_{0}=2.3980\times 10^{-4}$ mol.e$^2$.kcal$^{-1}$.\AA$^{-1}$) is vacuum permittivity, and $\mathrm{\Omega}$ is volume of the monolayer. The electric dipole moment was also calculated for a set of $N$ atoms with partial charges $q_i$ and positions ${\bf{r}}_i$ according to
\begin{align}
	\label{eq:diel2}
	{\bf{p}}=\sum_{i=1}^{N}(q_i-q_0){\bf{r}}_i,
\end{align}
where $q_0=\frac{1}{N}\sum_{i=1}^{N}q_i$ is the monopole component, subtracting from which renders Eq.~\ref{eq:diel2} independent of any specific choice of the origin.

Since in any computer simulation, no periodic system can infinitely be spread, cutting the monolayer then inevitably leads to non-integer total charge values based on the fact that both Si and N atoms have partial charges in making a periodic system. This is while applying MD simulations with full electrostatics requires the total charge to be vanishing; therefore, we could have tuned the charge on each atomic specie using the total charge neutrality condition: 
\begin{equation}
	q_{\mathrm{N}}N_{\mathrm{N}}+q_{\mathrm{Si}}N_{\mathrm{Si}}=0, 
\end{equation}
where $q_{\mathrm{i}}$ and $N_{\mathrm{i}}$ are the charge and number of each atomic specie, respectively. However, we did not do so, mainly because of its negligible contribution as the distance between the fibril and edges of the monolayer were large enough compared to dimensions of the fibril itself.

The figures were also rendered as follows: Fig.~\ref{fig:1} using XCrySDen~\cite{R46}; Figs.~\ref{fig:2},~\ref{fig:3}, and~\ref{fig:5} using VMD; Fig.~\ref{fig:4} using Gnuplot (version 5.2, patch-level 8)~\cite{R47}; and Figs.~\ref{fig:1},~\ref{fig:2},~\ref{fig:3}, and~\ref{fig:5} were also edited by Gimp 2.10.18~\cite{R48}.
\section{RESULTS}
Secondary structure of the fibril was found intact during 100 ns of the free-dynamics simulation of the reference system (A$\beta$-W). For A$\beta$-ML-W, Fig.~\ref{fig:3} shows the fibril at $t=0$ and $t=100$ ns on interaction with the Si$_3$N$_4$ monolayer. Comparing the two frames, the fibril's orientation with respect to monolayer has not changed, however, the A$\beta$-ML distance has decreased due to van der Waals interactions between fibril residues and the Si and N atoms of the monolayer, as observed in a number of works~\cite{q1,q2,q3,q4}.
\begin{widetext}
	\begin{minipage}{\linewidth}
\begin{figure}[H]
	\centering
	\subfigure[]{\label{subfig:3(a)}
	\includegraphics[scale=0.2]{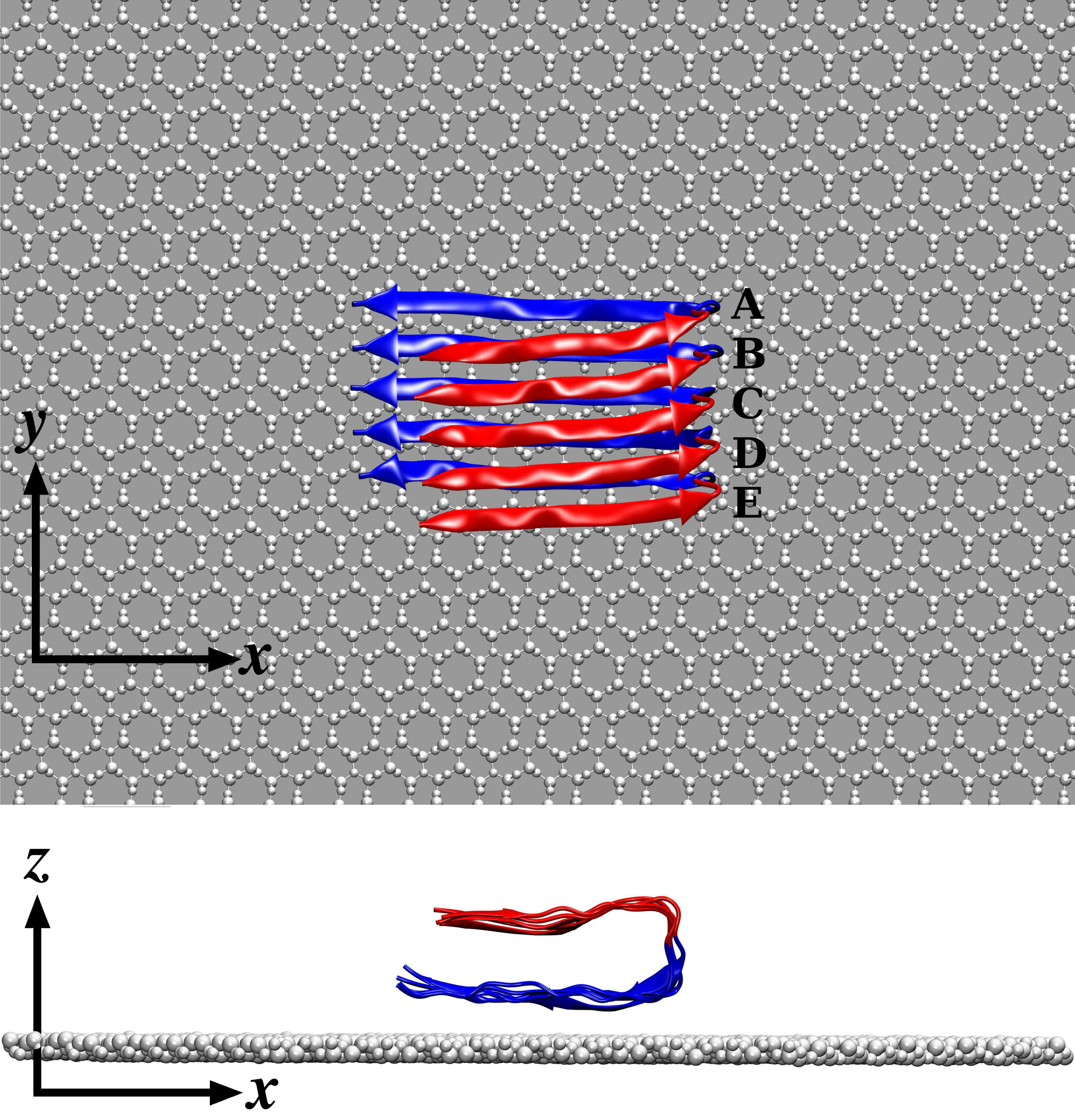}}
    \subfigure[]{\label{subfig:3(b)}
	\includegraphics[scale=0.2]{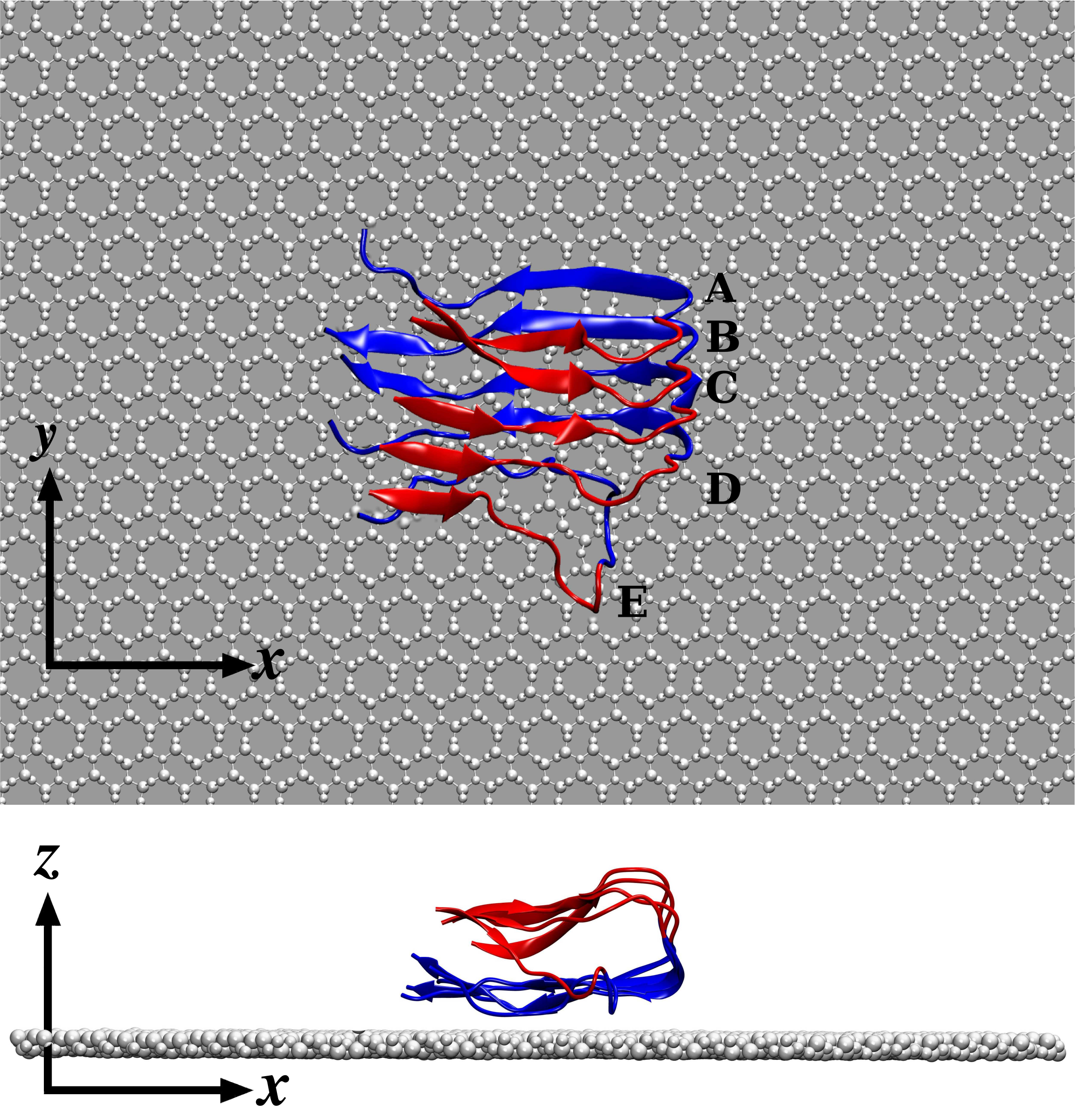}}
	\caption{\label{fig:3}
		Initial (a) and final (b) frames of the A$\beta$-ML-W simulation. The red and blue regions respectively represent the upper and lower $\beta$-sheets of the fibril; A to E are also chain labels. As is seen in (b), disintegration begins from E, and engages D over time.}
\end{figure}
\end{minipage}
\end{widetext}
Here, because N and Si have 3 and 4 empty electronic states, and form 2 and 3 covalent bonds with each other respectively (Fig.~\ref{fig:1}), either atomic specie is left with one empty state, accordingly making the monolayer capable of interacting with A$\beta$. 

The amino acid sequence KGAIIGLMVGGVVIA, numbered from 28 to 42, are closer to the monolayer due to electrostatic interaction with the monolayer, providing some sort of stability. 

From Fig.~\ref{subfig:3(b)}, fibril disintegration begins from chain E, also engaging D as time passes, showing a rather strong interaction between A$\beta$'s edge and ML. More investigation reveals that the amino acid sequence LVFFAEDVGSNK (numbered from 17 to 28) on E are responsible for such a behavior due to the fact that in the plane of the monolayer ($x-y$, Fig.~\ref{fig:1}), the two amino-acid sequences 17--28 and 28--42 are angled with respect to each other; therefore in chain E, the former (17--28) is not covered by the latter, being accordingly attracted to the monolayer. In contrast, the rest chains do not undergo such an attraction since the 17--28 sequences of A to D are covered by the 28--42 sequences of B to E, respectively. Moreover, the upper sequence LVFFAEDVGSNK from 17 to 28, albeit being farther from the monolayer compared to the lower one (28--42), interacts in a considerably stronger manner, initiating fibril disintegration as well. Indeed, the phenomenon that only one end deforms was already observed in the literature, the reason of which was also formerly revealed by Okumura {\textit{et al.}}~\cite{ittt}.

Fig.~\ref{fig:4} illustrates the time dependence of a number of MD indicators including root-mean-square deviation (RMSD), gyration radius, per-residue root-mean-square fluctuation (RMSF), number of hydrogen bonds, and SASA, both in presence and absence of the Si$_3$N$_4$ monolayer.
\begin{widetext}
	\begin{minipage}{\linewidth}
\begin{figure}[H]
	\centering
	\includegraphics[scale=0.8]{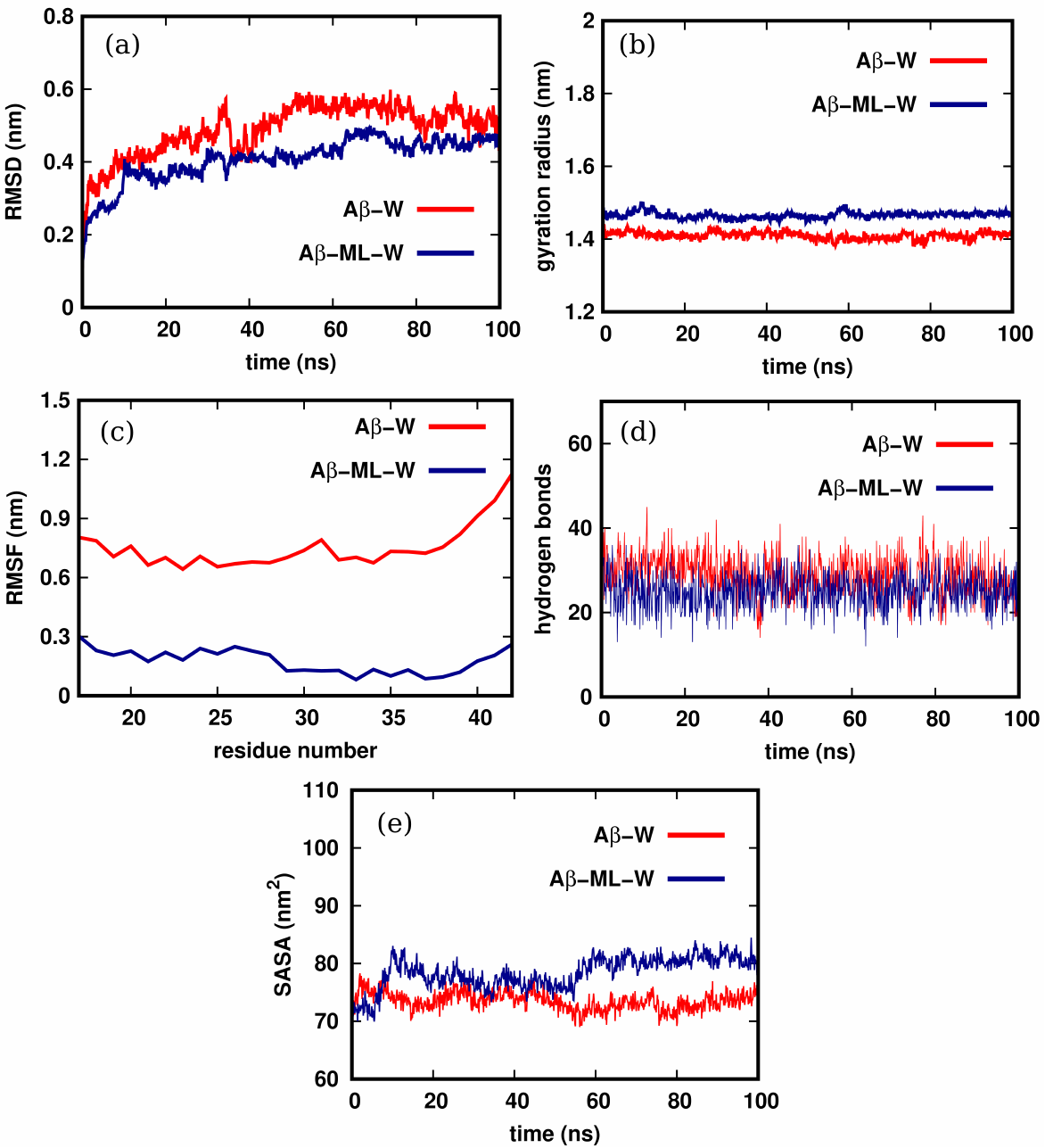}
	\caption{\label{fig:4}
		Conventional MD indicators including (a) RMSD, (b) radius of gyration, (c) per-residue RMSF, (d) number of hydrogen bonds, and (e) SASA calculated for A$\beta$ in absence (A$\beta$-W, red curve) and presence (A$\beta$-ML-W, blue curve) of the monolayer.}
\end{figure}
\end{minipage}
	\end{widetext}
From Fig.~\ref{fig:4}a, fibril's RMSD in A$\beta$-W is always larger than its analogue in A$\beta$-ML-W over the total time-span. Such a difference is mainly due to the fact that in the former, the fibril is free of any binding and therefore performs free dynamics in water, while in the latter, binding to the monolayer considerably affects its dynamics in a restrictive manner. Nevertheless, the blue curve has a totally increasing trend in contrast to the red one, in a way that they take nearly the same values at final time-steps, showing increase in fibril's dynamics due to fibril disintegration beginning from chain E (Fig.~\ref{subfig:3(b)}). 

According to Fig.~\ref{fig:4}b, gyration radius (compactness indicator) of the fibril on interaction with the monolayer is always larger than that in absence of the monolayer, meaning that the fibril in A$\beta$-ML-W gets less compact compared to A$\beta$-W as a chief consequence of fibril destabilization and disintegration. 

From Fig.~\ref{fig:4}c, RMSF of all the residues take considerably smaller values compared to those in absence of the monolayer obviously due to A$\beta$-ML binding. Reduction in the number of hydrogen bonds (Fig.~\ref{fig:4}d) on interaction with the Si$_3$N$_4$ monolayer compared to A$\beta$-W also shows decrease in the electrostatic (potential) energy between fibril's monomer peptides, leading to destabilization and change in secondary structure of the fibril as well.

Fibril disintegration also largely affects SASA as seen in Fig.~\ref{fig:4}e. Similarly, SASA in presence of the monolayer takes larger values compared to A$\beta$-W as a main consequence of the fact that the disintegrated fibril occupies a larger space, therefore, a wider surface area of the fibril could then be accessible to the solvent.
\begin{widetext}
	\begin{minipage}{\linewidth}
\begin{figure}[H]
	\centering
	\subfigure[]{\label{subfig:5(a)}
		\includegraphics[scale=0.555]{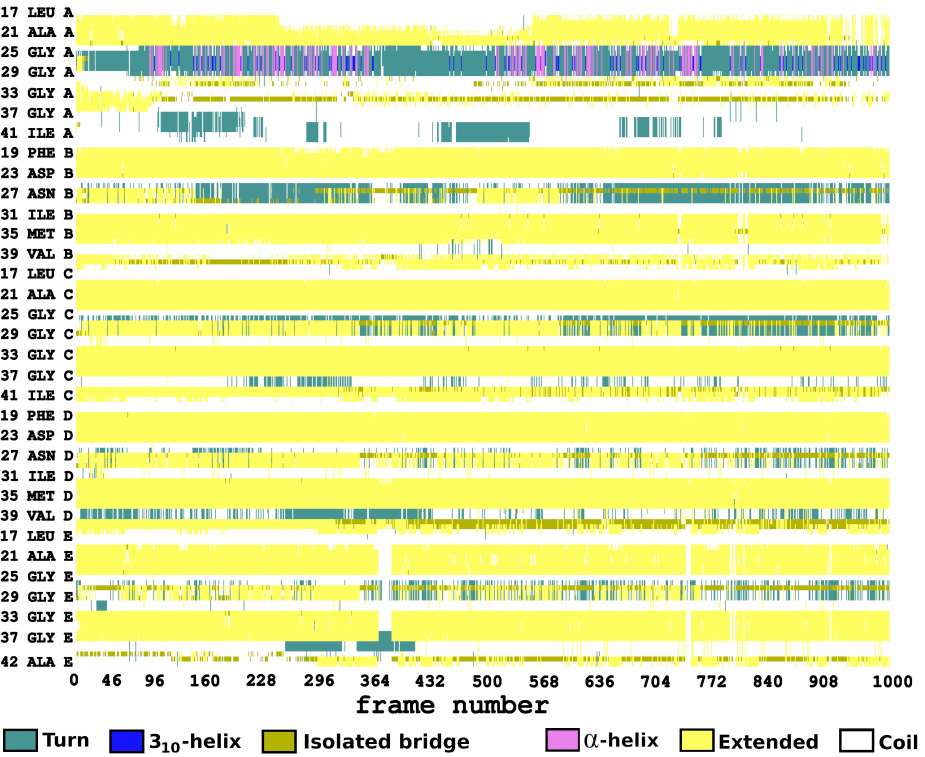}}
	\subfigure[]{\label{subfig:5(b)}
		\includegraphics[scale=0.555]{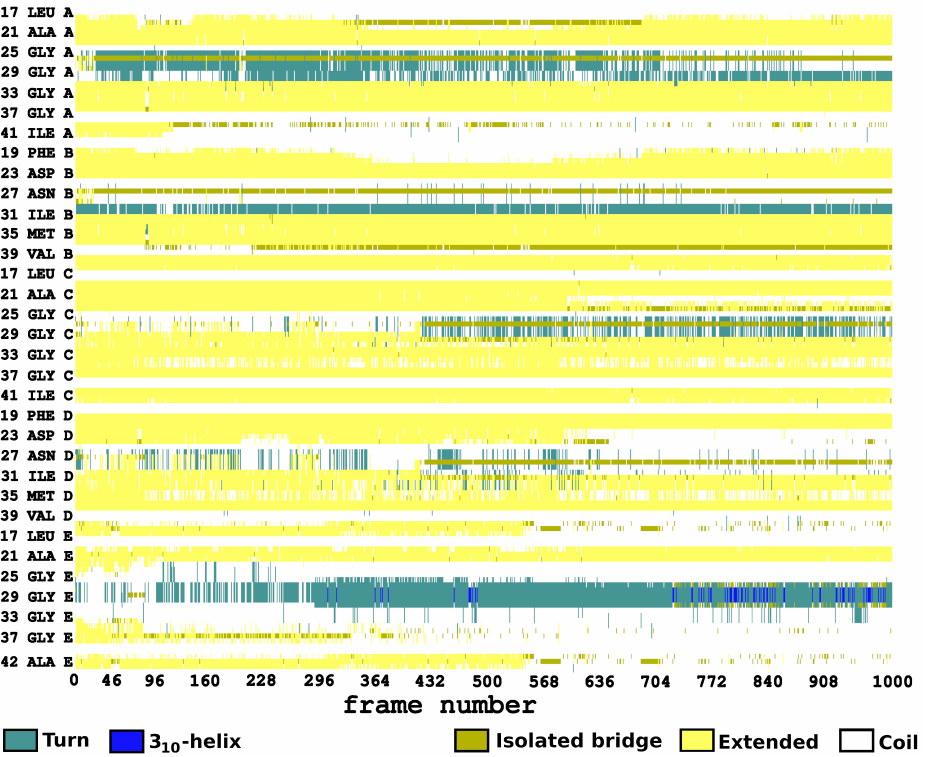}}
	\caption{\label{fig:5}
		Secondary structure of A$\beta$ fibril in (a) water (A$\beta$-W), and (b) on interaction with 2D-Si$_3$N$_4$ (A$\beta$-ML-W). A significant decrease in the $\beta$-sheet content at the place of chain E from 21 to 42 ALA can be seen. The legend shows the five secondary structures by different colors of which the fibril is composed. The vertical axis also indicates residue numbers (from 17 to 42 for each chain), the 3-letter residue names, and the corresponding chain labels to which the residues belong.}
\end{figure}
\end{minipage}
\end{widetext}
Fig.~\ref{fig:5} shows fibril's secondary structure as a function of frame number (or time). As is seen, either sub-figure is mostly yellow showing the dominance of the extended $\beta$-sheet content of the fibril in both A$\beta$-W and A$\beta$-ML-W. However, the $\beta$-sheet content is considerably lower in Fig.~\ref{subfig:5(b)} at chain E from 21 to 42 ALA due to fibril disintegration, turning the initial $\beta$-sheet-rich content into "turn" (green) and "coil" (white), and slightly into "isolated bridge" (light brown) at this chain. Such a finding, namely, disintegration at this place, has already been observed experimentally in an investigation using ultrasonic waves, in that the amino acids between 21 and 42 ALA were deformed first based on the fact that this region is formed by hydrophobic amino-acid residues, being attacked by a bubble generated by the ultrasonic wave~\cite{last}.
\section{\label{sec:5}CONCLUDING REMARKS}
Molecular dynamics (MD) simulations have been applied to investigate potential ability of semiconducting Si$_3$N$_4$ monolayer to disintegrate the structure/conformation of U-shaped amyloid beta (A$\beta$) fibrils. A number of conventional MD indicators have accordingly been calculated and analyzed including root-mean-square deviation, gyration radius, per-residue root-mean-square fluctuation, number of hydrogen bonds, and solvent-accessible surface area over the entire simulation trajectories. It has been found that, in agreement with the literature, disintegration initiates from the last chain (E) due to the rather strong interaction between the monolayer and the fibril's amino-acid sequence LVFFAEDVGSNK numbered from 17 to 28, making the next chain (D) to engage in disintegration as the time goes by. Accordingly, the $\beta$-sheet-rich content of chain E has considerably decreased on interaction with the monolayer, and turned into other secondary-structure types such as turn, coil, and isolated bridge, in accordance with experimental findings of the literature. Our results support the idea that Si$_3$N$_4$ monolayer has the potential of destabilizing the structure and conformation of U-shaped amyloid beta fibrils.\\


\begin{thebibliography}{}
\bibitem{R1}S.K. Mudedla, V. Subramanian, N.A. Murugan, H. Agren, Destabilization of amyloid ﬁbrils on interaction with MoS$_2$-based nanomaterials, RSC Adv. 9 (2019) 1613.
\bibitem{R2}R.E. Tanzi, L. Bertram, Twenty years of the Alzheimer's disease amyloid hypothesis: a genetic perspective, Cell 120 (2005) 545.
\bibitem{R3}R. Sultana, D.A. Butterfield, Alterations of some membrane transport proteins in Alzheimer's disease: role of amyloid $\beta$-peptide, Mol. BioSyst. 4 (2008) 36.
\bibitem{R4}J. Kang, H.G. Lemaire, A. Unterbeck, J.M. Salbaum, C.L. Masters, K.H. Grzeschik, G. Multhaup, K. Beyreuther, B. M{\"u}ller-Hill, The precursor of Alzheimer's disease amyloid A4 protein resembles a cell-surface receptor, Nature 325 (1987) 733.
\bibitem{R5}A.I. Bush, The metallobiology of Alzheimer's disease, Trends Neurosci. 26 (2003) 207.
\bibitem{R6}L.M. Young, P. Cao, D.P. Raleigh, A.E. Ashcroft, S.E. Radford, Ion Mobility Spectrometry--Mass Spectrometry Defines the Oligomeric Intermediates in Amylin Amyloid Formation and the Mode of Action of Inhibitors, J. Am. Chem. Soc. 136 (2014) 660.
\bibitem{R7}A.J. Doig, P. Derreumaux, Inhibition of protein aggregation and amyloid formation by small molecules, Curr. Opin. Struct. Biol. 30 (2015) 50.
\bibitem{R8}K. Debnath, S. Shekhar, V. Kumar, N.R. Jana, N.R. Jana, Efficient Inhibition of Protein Aggregation, Disintegration of Aggregates, and Lowering of Cytotoxicity by Green Tea Polyphenol-Based Self-Assembled Polymer Nanoparticles, ACS Appl. Mater. Interfaces 8 (2016) 20309.
\bibitem{R9}S.I. Yoo, M. Yang, J.R. Brender, V. Subramanian, K. Sun, N.E. Joo, S.-H. Jeong, A. Ramamoorthy, N.A. Kotov, Inhibition of amyloid peptide fibrillation by inorganic nanoparticles: functional similarities with proteins, Angew. Chem. 50 (2011) 5110.
\bibitem{R10}M. Richman, S. Wilk, N. Skirtenko, A. Perelman, S. Rahimipour, Surface-Modified Protein Microspheres Capture Amyloid-$\beta$ and Inhibit its Aggregation and Toxicity, Chem.--Eur. J. 17 (2011) 11171.
\bibitem{R11}S. Palmal, A.R. Maity, B.K. Singh, S. Basu, N.R. Jana, N.R. Jana, Inhibition of Amyloid Fibril Growth and Dissolution of Amyloid Fibrils by Curcumin--Gold Nanoparticles, Chem.--Eur. J. 20 (2014) 6184.
\bibitem{R12}S. Li, L. Wang, C.C. Chusuei, V.M. Suarez, P.L. Blackwelder, M. Micic, J. Orbulescu, R.M. Leblanc, Nontoxic carbon dots potently inhibit human insulin fibrillation, Chem. Mater. 27 (2015) 1764.
\bibitem{R13}J. Zhang, X. Zhou, Q. Yu, L. Yang, D. Sun, Y. Zhou, J. Liu, Epigallocatechin-3-gallate (EGCG)-stabilized selenium nanoparticles coated with Tet-1 peptide to reduce amyloid-$\beta$ aggregation and cytotoxicity, ACS Appl. Mater. Interfaces 6 (2014) 8475.
\bibitem{R14}M.J. Kogan, N.G. Bastus, R. Amigo, D. Grillo-Bosch, E. Araya, A. Turiel, A. Labarta, E. Giralt, V.F. Puntes, Nanoparticle-mediated local and remote manipulation of protein aggregation, Nano Lett. 6 (2006) 110.
\bibitem{R15}Q. Chen, L. Yang, C. Zheng, W. Zheng, J. Zhang, Y. Zhou, J. Liu, Mo polyoxometalate nanoclusters capable of inhibiting the aggregation of A$\beta$-peptide associated with Alzheimer's disease, Nanoscale 6 (2014) 6886.
\bibitem{R16}Z. Wang, C. Zhu, S.V. Bortolini, A. Hoffmann, H.X. Amari, Z.L. Liu, M.D. Dong, Dimensionality of carbon nanomaterial impacting on the modulation of amyloid peptide assembly, Nanotechnology 27 (2016) 304001.
\bibitem{R17}S. Bag, R. Mitra, S. Dasgupta, S.J. Dasgupta, Inhibition of Human Serum Albumin Fibrillation by Two-Dimensional Nanoparticles, J. Phys. Chem. B 121 (2017) 5474.
\bibitem{R18}J. Luo, S.K. W{\"a}rml{\"a}nder, C.H. Yu, K. Muhammad, A. Gr{\"a}slund, J.P. Abrahams, The A$\beta$ peptide forms non-amyloid fibrils in the presence of carbon nanotubes, Nanoscale 6 (2014) 6720.
\bibitem{R19}Y. Sun, Z. Qian, G. Wei, The inhibitory mechanism of a fullerene derivative against amyloid-$\beta$ peptide aggregation: an atomistic simulation study, Phys. Chem. Chem. Phys. 18 (2016) 12582.
\bibitem{R20}Z. Yang, C. Ge, J. Liu, Y. Chong, Z. Gu, C.A. Jimenez-Cruz, Z. Chaia, R. Zhou, Destruction of amyloid fibrils by graphene through penetration and extraction of peptides, Nanoscale 7 (2015) 18725.
\bibitem{R22}A. Shekaari, M.R. Abolhassani, First-principles investigation of the thermodynamic properties of two-dimensional MoS$_2$, Chin. J. Phys. 55 (2017) 105.
\bibitem{R24}A. Shekaari, M. Jafari, Unveiling the first post-graphene member of silicon nitrides: A novel 2D material, Comput. Mater. Sci. 180 (2020) 109693.
\bibitem{R26}M. Li, A. Zhao, K. Dong, W. Li, J. Ren, X. Qu, Chemically exfoliated WS$_2$ nanosheets efficiently inhibit amyloid $\beta$-peptide aggregation and can be used for photothermal treatment of Alzheimer's disease, Nano Res. 8 (2015) 3216.
\bibitem{R27}A. Shekaari, M. Jafari, Biocompatibility of 2D silicon nitride: interaction at the nano-bio interface, Mater. Res. Express 8 (2021) 095404.
\bibitem{h}J.M. Haile, Molecular Dynamics Simulation: Elementary Methods, Wiley, New York, 1992.
\bibitem{man}A. Shekaari, M. Jafari, arxiv:2103.16944.
\bibitem{R28}T. Luhrs, C. Ritter, M. Adrian, D. Riek-Loher, B. Bohrmann, H. Dobeli, D. Schubert, R. Riek, 3D structure of Alzheimer's amyloid-$\beta$(1--42) fibrils, PNAS 102 (2005) 17342.
\bibitem{R29}J.C. Phillips, R. Braun, W. Wang, J. Gumbart, E. Tajkhorshid, E. Villa, C. Chipot, R.D. Skeel, L. Kale, K. Schulten, Scalable molecular dynamics with NAMD, J. Comput. Chem. 26 (2005) 1781.
\bibitem{R30}L. Torvalds, The Linux edge, Commun. ACM 42 (1999) 38.
\bibitem{R31}K. Vanommeslaeghe, E. Hatcher, C. Acharya, S. Kundu, S. Zhong, J. Shim, E. Darian, O. Guvench, P. Lopes, I. Vorobyov, A.D. Mackerell Jr., CHARMM general force field: A force field for drug-like molecules compatible with the CHARMM all-atom additive biological force fields, J. Comput. Chem. 31 (2010) 671.
\bibitem{R32}W.L. Jorgensen, J. Chandrasekhar, J.D. Madura, R.W. Impey, M.L. Klein, Comparison of simple potential functions for simulating liquid water, J. Chem. Phys. 79 (1983) 926.
\bibitem{R33}W. Humphrey, A. Dalke, K. Schulten, VMD: Visual molecular dynamics, J. Mol. Graphics 14 (1996) 33.
\bibitem{R34}T. Darden, D. York, L. Pedersen, Particle mesh Ewald: An $N.\log(N)$ method for Ewald sums in large systems, J. Chem. Phys. 98 (1993) 10089.
\bibitem{R35}B. Lee, F.M. Richards, The interpretation of protein structures: estimation of static accessibility, J. Mol. Biol. 55 (1971) 379.
\bibitem{R36}A. Shrake, J.A. Rupley, Environment and exposure to solvent of protein atoms. Lysozyme and insulin, J. Mol. Biol. 79 (1973) 351.
\bibitem{R37}J.A. Wendel, W.A. Goddard, The Hessian biased force field for silicon nitride ceramics: Predictions of thermodynamic and mechanical properties for $\alpha-$ and $\beta-$Si$_3$N$_4$, J. Chem. Phys. 97 (1992) 5048.
\bibitem{R39}J.P. Perdew, K. Burke, M. Ernzerhof, Generalized Gradient Approximation Made Simple, Phys. Rev. Lett. 77 (1996) 3865.
\bibitem{R40}R.G. Parr, W. Yang, Density-Functional Theory of Atoms and Molecules, Oxford University Press, Oxford, New York, 1989.
\bibitem{R38}P. Giannozzi, S. Baroni, N. Bonini, M. Calandra, R. Car, C. Cavazzoni, D. Ceresoli, G.L. Chiarotti, M. Cococcioni, I. Dabo, A.D. Corso, S. de Gironcoli, S. Fabris, G. Fratesi, R. Gebauer, U. Gerstmann, C. Gougoussis, A. Kokalj, M. Lazzeri, L. Martin-Samos, N. Marzari, F. Mauri, R. Mazzarello, S. Paolini, A. Pasquarello, L. Paulatto, C. Sbraccia, S. Scandolo, G. Sclauzero, A.P. Seitsonen, A. Smogunov, P. Umari, R.M. Wentzcovitch, \qe: a modular and open-source software project for quantum simulations of materials, J. Phys. Condens. Matter 21 (2009) 395502.
\bibitem{R41}A.M. Rappe, K.M. Rabe, E. Kaxiras, J.D. Joannopoulos, Optimized pseudopotentials, Phys. Rev. B 41 (1990) 1227.
\bibitem{R42}S.G. Louie, S. Froyen, M.L. Cohen, Nonlinear ionic pseudopotentials in spin-density-functional calculations, Phys. Rev. B 26 (1982) 1738.
\bibitem{R43}F.D. Murnaghan, The Compressibility of Media under Extreme Pressures, PNAS 30 (1944) 244.
\bibitem{R44}J. Kohanoff, Electronic Structure Calculations for Solids and Molecules: Theory and Computational Methods, Cambridge University Press, Cambridge, 2006.
\bibitem{R45}F. Tassone, F. Mauri, R. Car, Acceleration schemes for {\em{ab initio}} molecular-dynamics simulations and electronic-structure calculations, Phys. Rev. B 50 (1994) 10561.
\bibitem{R46}A. Kokalj, XCrySDen--a new program for displaying crystalline structures and electron densities, J. Mol. Graph. Model. 17 (1999) 176.
\bibitem{R47}www.gnuplot.info.
\bibitem{R48}https://www.gimp.org.
\bibitem{q1}M. Serra-Batiste, M. Ninot-Pedrosa, M. Bayoumi, M. Gair{\'i}, G. Maglia, N. Carulla, A$\beta$42 assembles into specific $\beta$-barrel pore-forming oligomers in membrane-mimicking environments, Proc. Natl. Acad. Sci. U.S.A. 113 (2016) 10866.
\bibitem{q2}S.G. Itoh, M. Yagi-Utsumi, K. Kato, H. Okumura, Effects of a hydrophilic/hydrophobic interface on amyloid-$\beta$ peptides studied by molecular dynamics simulations and NMR experiments, J. Phys. Chem. B 123 (2019) 160.
\bibitem{q3}H. Okumura, S.G. Itoh, Molecular dynamics simulations of amyloid-$\beta$(16-22) peptide aggregation at air-water interfaces, J. Chem. Phys. 151 (2020) 095101.
\bibitem{q4}N. Agrawal, A.A. Skelton, E. Parisini, A coarse-grained molecular dynamics investigation on spontaneous binding of A$\beta_{1-40}$ fibrils with cholesterol-mixed DPPC bilayers, Comput. Struct. Biotechnol. J. 21 (2023) 2688.
\bibitem{ittt}H. Okumura, S.G. Itoh, Structural and fluctuational difference between two ends of A$\beta$ amyloid fibril: MD simulations predict only one end has open conformations, Sci. Rep. 6 (2016) 38422.
\bibitem{last}H. Okumura, S.G. Itoh, Amyloid fibril disruption by ultrasonic cavitation: Nonequilibrium molecular dynamics simulations, J. Am. Chem. Soc. 136 (2014) 10549.
\end{thebibliography}
\end{document}